# Studies of self-organized Nanostructures on InP(111) surfaces after low energy Ar$^+$ ion irradiation

D. Paramanik, S. Majumdar, S.R. Sahoo, S.N. Sahu and S. Varma*

Institute of Physics, Bhubaneswar - 751005, India

**Abstract**

We report formation of self organized InP nano dots using 3 keV Ar$^+$ ion sputtering, at 15° incidence from surface normal, on InP(111) surface. Morphology and optical properties of the sputtered surface, as a function of sputtering time, have been investigated by Scanning Probe Microscopy and Raman Scattering techniques. Uniform patterns of nano dots are observed for different durations of sputtering. The sizes and the heights of these nano dots vary between 10 to 100 nm and 20 to 40 nm, respectively. With increasing of sputtering time, t, the size and height of these nano dots increases up to a certain sputtering time $t_c$. However beyond $t_c$, the dots break down into smaller nanostructures, and as a result, the size and height of these nanostructures decrease. The uniformity and regularity of these structures are also lost for sputtering beyond $t_c$. The crossover behavior is also observed in the rms surface roughness. Raman investigations of InP nano dots reveal optical phonon softening due to phonon confinement in the surface nano dots

*Corresponding author: shikha@iopb.res.in

# INTRODUCTION

With the fast growing interest in nanotechnology, fabrication of regular semiconductor nanostructures with controlled size and height is of great importance. InP is a vastly used semiconductor material in this area. Ion beam sputtering is frequently regarded as an alternative process for the fabrication of various nano-structured surfaces or interfaces via self organization. Under certain conditions sputtering can roughen this surface resulting in a pronounced topography evolution in some cases producing well ordered pattern. This pattern formation is related to the surface instability between curvature dependent ion sputtering that roughens the surface and smoothing by different relaxation mechanism [1,2]. Recent studies show the formation of ordered InP nanostructure by Ar+ ion sputtering under normal ion incidence [3-5] or alternatively under oblique ion incidence with simultaneous sample rotation [6]. Generally for off-normal ion incidence without sample rotation, a periodic height modulation in the form of ripple or wave like structure with a sub-micron length scale develops during ion bombardment as observed for single crystalline III-V semiconductors [7,8].

Starting from the Bradley-Harper (BH) theory [2] a successful description of the morphological evolution of the ion sputtered surface is described by the isotropic Kuramoto-Sivashinsky (KS) equation [4,5,9,10] the temporal development of the surface profile h(x,y,t) is given by the following undamped KS equation:

$$\frac{\partial h}{\partial t} = -V_0 + \nu \nabla^2 h - D_{eff} \nabla^4 h + \frac{\lambda}{2}(\nabla h)^2 + \eta \qquad (1)$$

where $V_0$ is the constant erosion velocity and $\nu$ is the effective surface tension, caused by the erosion process, which usually has a negative value leading to a surface instability. $D_{eff}$ is the surface diffusion coefficient which is the sum of thermal diffusion and ion induced effective diffusion. The nonlinear term $\lambda/2(\nabla h)^2$ accounts for the slope-dependent erosion yield that brings forth the saturation of surface roughness with time. $\eta$ is an uncorrelated white noise with zero mean, mimicking the randomness resulting from the stochastic nature of ion arrival to the surface.

In contrast, arrays of zero or two dimensional nanostructure can be produced by ion sputtering under normal or oblique (> 30° from normal) ion incidence geometry with simultaneous sample rotation[5,6]. Here we report for the first time the formation of self organized InP nanostructure for slight off-normal (15° from normal) ion incidence by 3keV Ar ion sputtering. In this letter we also report the Scanning Probe Microscopy (SPM) studies of the systematic changes in size and height of the nanostructures with increasing sputtering time. We have produced these nanostructures over a wide temporal range spanning from 5 min to 80 min, where the fluence for one minute sputtering is calculated to be $1.3 \times 10^{16}$ ions/cm$^2$. The average size and height of the nanostructures varies in the range of 10 to 100 nm and 5 to 40 nm, respectively, for different sputtering times. The nanostructures develop in the early stages of sputtering and grow in size and height with increasing time upto a certain time. We find a crossover time $t_c$ upto which the size and the height of the nanostructure increases, however beyond $t_c$ they start to decrease. The value of rms surface also increases upto $t_c$ and it decreases for longer sputtering durations. Raman spectroscopy has been utilized to study the structural modification of the sputtered surface.

## EXPERIMENT

Samples used in this work are commercially available epi-polished InP(111) wafers. Sputtering was preformed in a ultra high vacuum chamber with a base pressure of $1\times10^{-9}$ torr. InP(111) wafers were placed on the sample holder and irradiated with $Ar^+$ beam from an EX05 ion gun from VG Microtech. The beam with 3 keV energy was utilized to irradiate the sample surface. The angle of irradiation was 15° from normal to surface. The beam was focused on a circular spot of diameter 0.3 cm on the sample surface with ion current density 10 µA/cm$^2$. Samples were sputtered for 5, 10, 20, 30, 40, 60 and 80 minutes without any sample rotation. The fluence for one minute of sputtering is calculated to be $3\times10^{16}$ ions/cm$^2$. The bombarded samples were studied in tapping mode in a Veeco Nanoscope IIIA multimode Scanning Probe Microscope (SPM). Several images of scan length 50 nm to 10 µm were acquired and analyzed. Raman scattering measurements were performed using a SPEX 1877E Triplemate Spectrometer with a liquid nitrogen cooled, charged coupled device array. The laser power was kept low, at 500 mW, to avoid laser annealing effect on the sample. Raman experiments were carried out at room temperature using the 488 nm line of an argon ion laser in the backscattering geometry.

## RESULTS AND DISCUSSION

Figure 1 shows the two dimensional 500 nm×500 nm AFM images of the surfaces that are sputtered for different times from 10 minutes to 80 minutes. For comparison, the virgin(unsputtered) sample is also shown here. Fig.2a and 2b show the average size and height distributions, respectively, of the nanostructures for different sputtering time from 10 minutes to 80 minutes. In the early stage of sputtering i.e. sputtering for 5 min, the surface is dominated by small, wavy perturbation generated by the interplay between the ion induced instability and surface relaxation [4]. After sputtering for 10 minutes, these structures develop into isolated nano dots(fig.1a). A highly uniform and narrow distribution of dots with an average size of 25 nm and height of 4 nm is observed . The density of dots is found to be $2\times10^{11}$ cm$^{-2}$. For 20 minutes of sputtering, although the density of nanostructure does not change, the average size slightly increases and the structures start to overlap each other. The average size and height of the nanostructures is 32 nm and 8 nm, respectively (fig.2). After 30 minute of sputtering several small nanostructures join together and form big nanostructures. The average size and height of these nanostructure is 45 nm and 18 nm, respectively. Since several nanostructures join and form bigger nanostructures, the density of nanostructure decreases to $4\times10^{10}$ cm$^{-2}$ . The process of ripening and agglomeration of nanostructures continues upto 40 minutes of sputtering. At 40 minutes, most uniformly distributed and regular shaped nanostructure are observed (fig.1b). Here the nanostructure appear to be of rectangular shape with roughly uniform orientation. Figures 2a and 2b show immense increase in the size and the height of the dots at this stage with their average size and height becoming 90 nm and 42 nm, respectively. The density of the dots is observed to be $6\times10^9$ cm$^{-2}$. Further sputtering beyond 40 minutes causes breaking down of these big structures into smaller sized nanostructures. The nanostructures formed at 60 min sputtering are smaller in size and lower in height. At this stage, the average size and height are 78 nm and 25 nm, respectively. Further increase in the sputtering time to 80 minutes causes more randomness and decrease in size and lowering in height on nanostructures. This causes an increase in their density to $5\times10^{10}$ cm$^{-2}$. The structures formed at 80 min sputtering have average size of 42 nm and height of 14 nm.

Above study of nanostructure formation and distribution of their size and height for different sputtering time reveals that there is a crossover time $t_c$ which is 40 minutes where the most regular and uniform nanostructures are formed. However for sputtering beyond $t_c$, the size and height of the nanostructures decreases and the uniformity and regularity of their distributions are destroyed (fig.1 and 2). Using the KS model, Kahng et al [4] theoretically proved that there is crossover sputtering time $t_c$, where most uniform nanostructure are formed. For sputtering upto $t_c$ the size and height of the nanostructure increases and at $t_c$ they become maximum.

The crossover behavior is also observed through the rms surface roughness. Fig.3 shows rms roughness of the InP surface for sputtering from 5 minutes to 80 minutes. This quantity exhibits a sharp transition at 40 minutes, which is the crossover time in this study. For $t<t_c$, rms roughness increases from about 0.5 nm for the virgin sample to 6.5 nm for 40 min of sputtering. For $t>t_c$ surface roughness decreases to about 2.9 nm for 80 min of sputtering. Several studies[4, 6]on InP surfaces demonstrate a saturation of surface roughness for long sputtering durations. However, none of these studies displayed any surface smoothening behavior.

Fig.4 shows the room temperature Raman spectra of virgin (bulk) InP and nano patterned InP surface after Ar$^+$ ion sputtering for times varying from 5 minutes to 40 minutes. The Raman spectrum of virgin InP(111) shows both longitudinal optical (LO) and transverse optical (TO) modes at 351.5 cm$^{-1}$ and 301.5 cm$^{-1}$, respectively. The broad phonon peak at 306.0 cm$^{-1}$ can be assigned as the L$^-$ LO phonon plasmon coupled (LOPC) mode in InP. This coupled mode has been reported earlier by us for MeV ion implanted InP(111) samples for different fluences[11]. The Raman spectrum after 5 min sputtering shows shifts for both the LO and TO phonons towards lower wave numbers. For this sample, the LO and TO phonons modes appear at 350.5 and 300.5 cm$^{-1}$, respectively. The peaks also appear asymmetric. After 10 min and 20min sputtering the shifts in the LO and TO modes are small. The LO and TO modes for 10min sputtering appear at 350.2 and 300.3 cm$^{-1}$, respectively, and for 20min sputtering at 350.3 and 300.4 cm$^{-1}$, respectively. LO and TO modes display large shifts compared to virgin, 2.3 and 1.7 cm$^{-1}$, respectively, after 30 of sputtering with the modes appearing at 349.2 and 299.8 cm$^{-1}$, respectively. For 40min of sputtering the LO and TO modes are observed at 349.7 and 299.9 cm$^{-1}$, respectively. The shift in the LO and TO modes as well as the asymmetricity in these peaks after 30min, appear more than that seen after 40min. All this may be due to the higher phonon confinement due to smaller InP nanodots [11] at 30min.

Thus, it is clear that with the change in size and height of the nano dots for different sputtering, there are different amount of shifting of both LO and TO phonons as a result of phonon confinement effect. This shifting is due to the phonon confinement in nano dots of varying size and height distributions shown in figures 2a and 2b. It is assumed that such a low energy sputtering may not cause lattice amorphicity. However, the Raman line shape will also be controlled by the presence of lattice strain and disorder in the sample after sputtering.

Correlating these results with the observed surface morphologies, we find that the irregular patterns are induced on InP surface during the early 5 minutes of sputtering process. The surface, at this stage, is dominated by small, wavy perturbations generated by the interplay between the ion induced instability and the surface relaxation [4]. The amplification of the random amplitudes by the negative surface tension compete with the smoothening processes such as surface diffusion and viscous flow. This leads to the formation of regular surface patterns after 10 min of sputtering. During early stages of sputtering, upto $t_c$ = 40 min, the dots grow in size and agglomerate. This evolution is also accompanied by surface roughening. The surface at $t_c$ is characterized by the presence of rectangular shaped nano dots. For sputtering beyond $t_c$, the nanodots reduce in size and height and a reduction in the uniformity of nanostructures is also observed.

Based on the results presented in fig 1, 2 and 3 the following scenario emerges for QD formation. In the early stages (for 5, 10, 20, 30 minutes) of the erosion process, we observe the formation and growth of nanostructure upto $t_c$ (40 min). However as time increases beyond $t_c$ the nonlinear term leads to breaking up of dots. As a result, bigger structures break down into multiple smaller structures and the uniformity of the distribution is lost. The most uniform distribution with the maximum size and height of the nanostructures appear at the characteristic time $t_c$. This time depends on the flux and the ion energy, with the characteristic size depending on the ripple period [4]. As time increases beyond $t_c$, the nonlinear term leads to kinetic roughening of the surface at large length scales [12,13] and while the QD do not disappear they break into smaller nanodots.

## SUMMARY AND CONCLUSION

In summary, we have observed that 3 keV $Ar^+$ ion sputtering of InP(111), at 15° incidence, causes the formation nano dots. The size and height distributions of these nano dots depend on the sputtering time. We find a crossover time $t_c$ where the size and height of the nano dots become maximum. Similar behavior of cross over time is also observed in rms roughness. The Raman investigation indicates both LO and TO phonon softening on InP(111) surface. This is due to the effect of phonon confinement in nano dots.

## ACKNOWLEDGMENTS

This work is partly supported by ONR grant no. N00014-97-1-0991. We would like to acknowledge the help of S.K. Choudhury during sputtering and we would like to thank S.N. Sarangi for Raman scattering measurements.

**FIGURES**

Fig.1: 500 nm× 500 nm AFM images of InP surfaces for the virgin sample as well as after sputtering with 3 keV Ar$^+$ ions for a time of (a) 10 min, (b) 40 min, and (c) 80 minutes. The color height scale bar with different Z values is also shown in the picture.

Fig.2: Variation of average size and height of the nano dots with sputtering time.

Fig.3: Variation of rms surface roughness with sputtering time.

Fig.4: Raman spectra of virgin InP(111) and 3 keV Ar ion sputtered InP(111) for time 5 min (a), 10 min (b), 20min (c), 30 min (d), and 40 min (e).

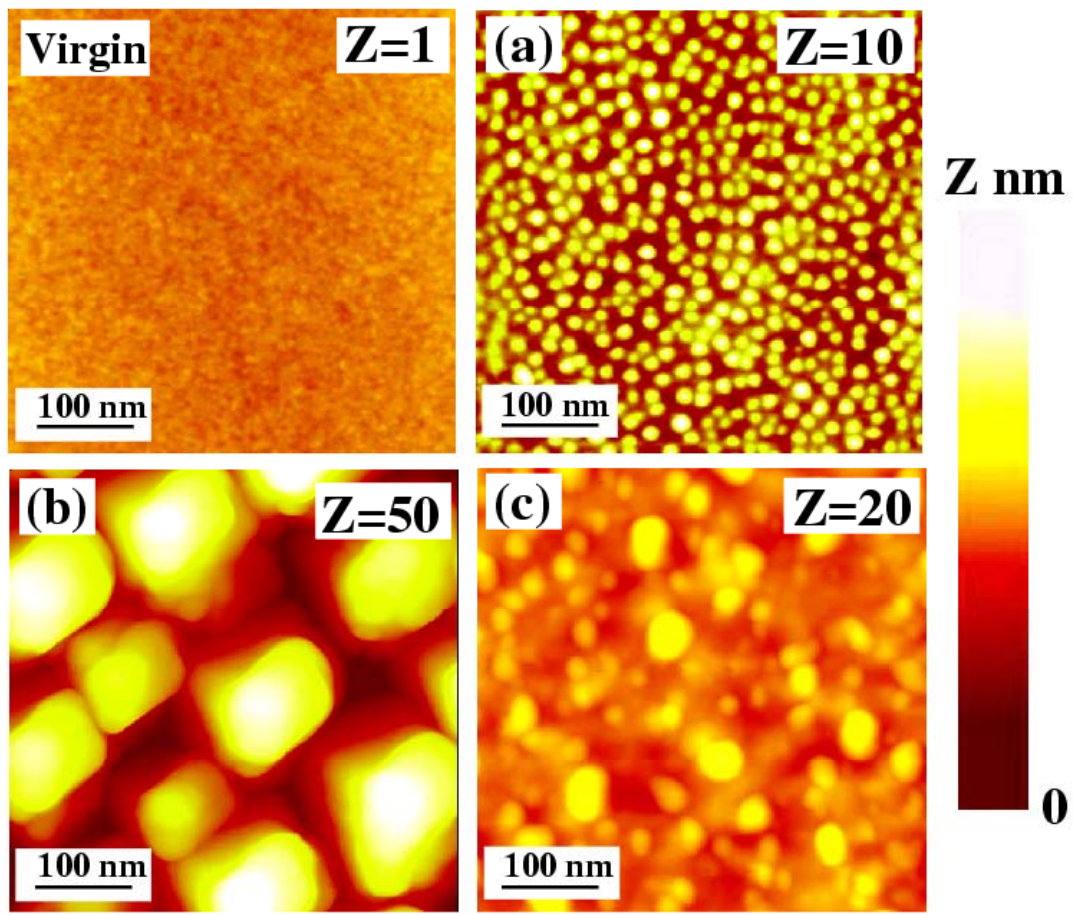

**Figure 1:
Paramanik et al.**

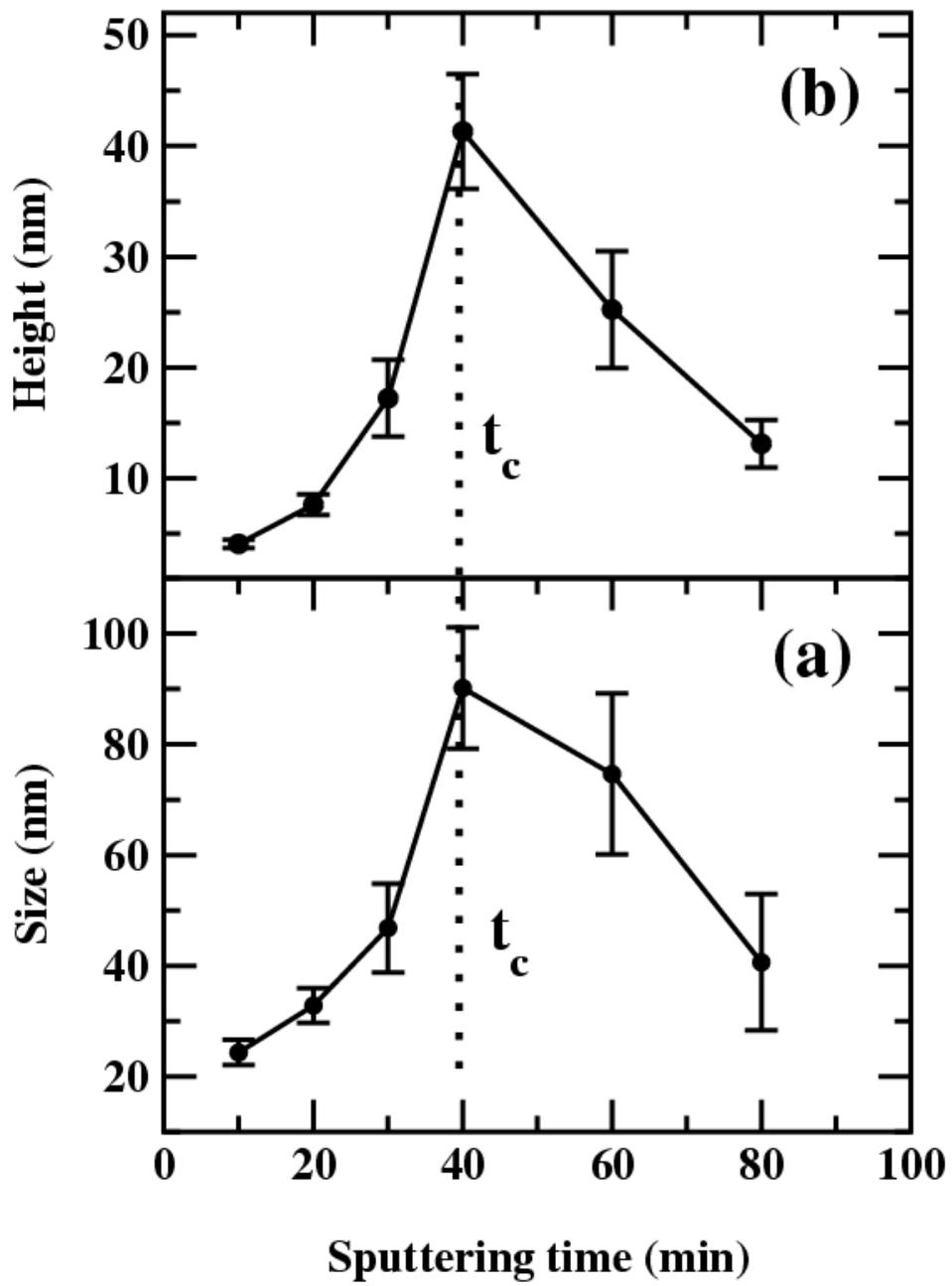

**Figure 2:
Paramanik et al.**

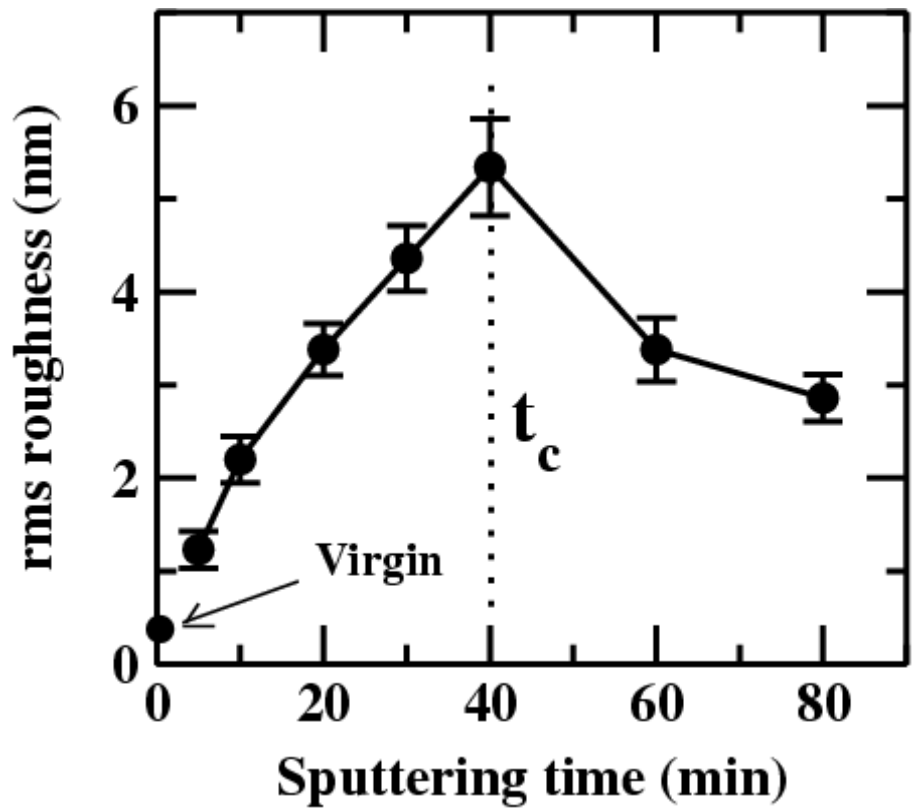

**Figure 3:**
**Paramanik et al.**

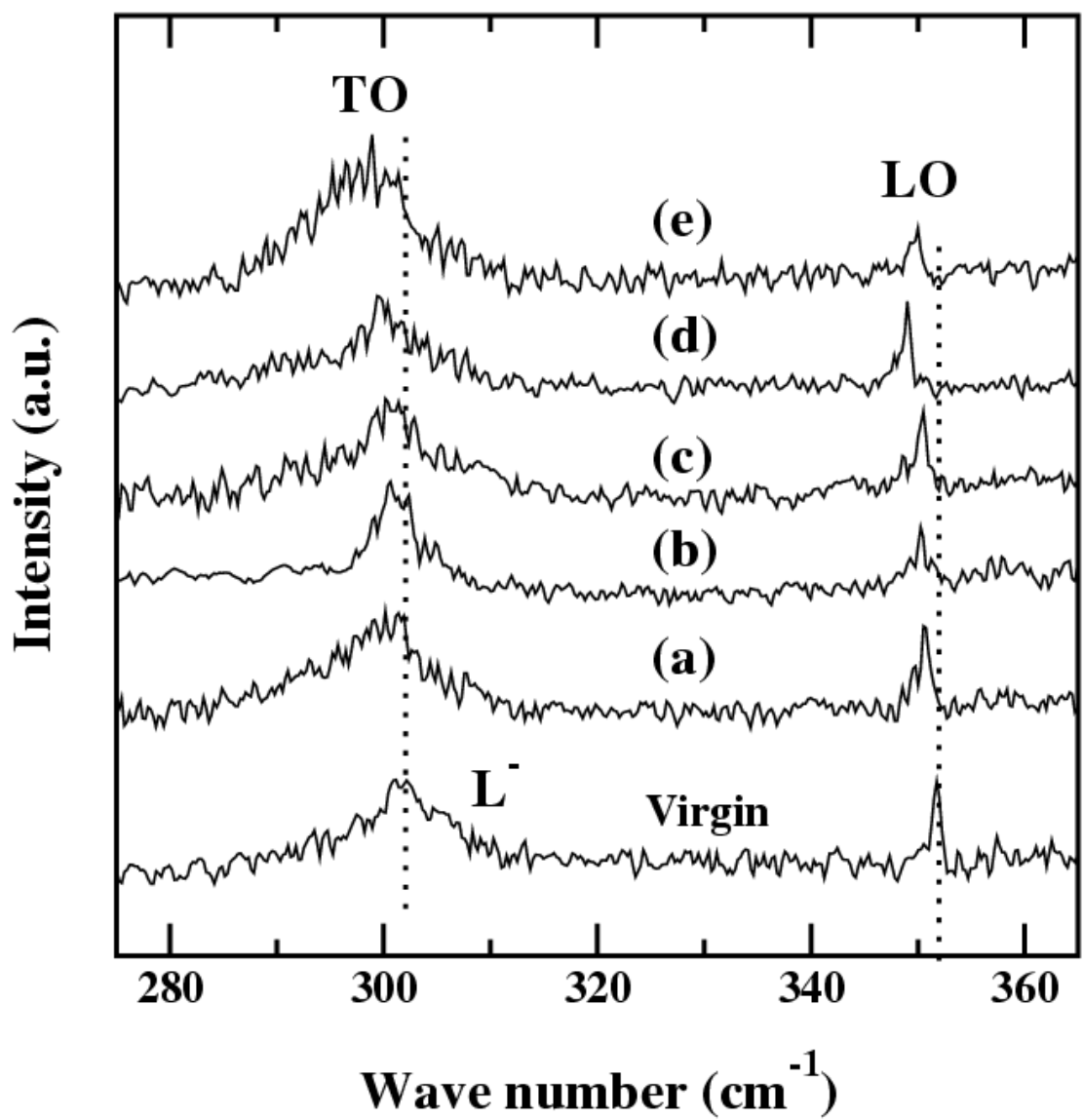

**Figure 4:
Paramanik et al.**